\documentclass[showkeys]{revtex4-2}

\usepackage{multirow}
\usepackage{orcidlink}

\begin{document}


\title{Measurement of the O$^{-}$ Photodetachment cross-section in the electrostatic storage ring FLSR}

\author{Oliver Forstner$^{1,2,3,*}$\orcidlink{0000-0003-3636-4669}, Thorben Niemeyer$^{4}$, Andrey I. Bondarev$^{2,3}$, Markus Dworak$^{5}$, Vadim Gadelshin$^{4}$, Raphael Hasse$^{4}$, Lothar Schmidt$^{5}$, Markus Schöffler$^{5}$, Matou Stemmler$^{4}$, Kurt E. Stiebing$^{5}$, Dominik Studer$^{4}$, Klaus Wendt$^{4}$, Patric Ziel$^{5}$}

\affiliation{$^1$Institut für Optik und Quantenelektronik, Friedrich-Schiller-Universität, D-07743 Jena, Germany}

\affiliation{$^2$Helmholtz Institute Jena, D-07743 Jena, Germany}

\affiliation{$^3$GSI Helmholtzzentrum für Schwerionenforschung GmbH, D-64291 Darmstadt, Germany}

\affiliation{$^4$Institute of Physics, Johannes Gutenberg University, D-55128 Mainz, Germany}

\affiliation{$^5$Institute of Nuclear Physics, Johann Wolfgang Goethe-Universität, D-60438 Frankfurt am Main, Germany}

\affiliation{$^*$Author to whom any correspondence should be addressed.}

\email{oliver.forstner@uni-jena.de}


\begin{abstract}
The achievable precision of photodetachment studies via tunable laser light has been investigated on circulating oxygen anions at the room-temperature electrostatic low-energy storage ring of Frankfurt University (FLSR) in preparation of  investigations on molecular species. For this purpose, the FLSR, originally designed for reaction dynamics studies on positive ions, was upgraded by a source for anions, by the installation of a tunable high repetition rate laser system and by implementation of a transversal interaction region between laser and ion beam. Results on the electron affinity and the ground state fine structure of O$^{-}$, which both serve as a calibration standard, are discussed, exhibiting the performance and accuracy of this new spectroscopic application of the FLSR.
\end{abstract}

\maketitle

\section{Introduction}
Negative ions are fragile quantum systems in which an additional electron is bound to a neutral atom or molecule. Let us focus here on anions of atoms only. A polarization of the electron shell of the neutral atom by the influence of the surplus electron may result in a weak attractive Van-der-Waals like potential with $1/r^{4}$ radial dependence \cite{And04}. This leads to the situation, that more than 80\,\% of all elements of the Periodic Table are able to form stable negative ions. On the other hand, due to the lack of the strong and long-range Coulomb force only in five exceptional cases of anions, namely La$^{-}$, Ce$^{-}$, Os$^{-}$, Th$^{-}$ and U$^{-}$, excited states with opposite parity are present or have been discovered so far in the level scheme of a specific anion \cite{Bil00,Wal11,Kel14,Wal14,Tan19,Tan21}. Hence, most negative ions are inaccessible to conventional laser spectroscopy techniques apart of studying photodetachment processes.

The most relevant and easily accessible property, which is characteristic for a specific anion in the gas phase, wherein it experiences no mutual disturbances as would be the case in the other phases, is the binding energy of the additional electron, defined as the positive-valued electron affinity (EA), whereas a negative value corresponds to an unbound system. Experimentally, the EA value is easily accessible and conventionally measured via removal of the excess electron, employing either photodetachment processes based on electromagnetic radiation well-tuned to the energy required for neutralization or energy-resolved charge-exchange in collisions processes. Both techniques are usually applied in dedicated mass spectrometric instruments or storage devices, specifically adapted for handling of the delicate systems of anions. A well-suitable overview of these techniques is given in the review article by Hotop and Lineberger \cite{Hot75}. Today, the EAs of more than 70 elements have been measured experimentally and are tabulated \cite{And99}. Specifically, laser photodetachment spectroscopy in the visible to infra-red range using tunable high-power lasers serves as a high-resolution technique to study neutralization threshold structures of anions. This technique has delivered the EA data of highest precision and opened the path to concisely study substructures, which are caused, e.g., by fine structure splitting of the ground states of either the negative ion or the generated atom. Precisions achieved are typically in the order of 10$^{-5}$ up to 10$^{-7}$\,eV, significantly decreasing for anions of higher atomic number Z.

In particular, electron affinity studies in the element oxygen have played an outstanding role in the development of these high precision measurements on anions. Diverse sets of experimental data were determined experimentally multiple times during the last 50 years, with different techniques and frequently not in full agreement, as discussed, e.g., in \cite{Blo95} and two review papers \cite{And04,And99}. In addition, numerous theoretical calculations have been attempted, unfortunately not yet matching the experimentally achieved accuracy \cite{God99}. O$^{-}$ exhibits a pure p-electron detachment, which should perfectly follow the Wigner law near threshold but is characterized by fine structures in the anionic and neutral states. Today the EA is known with outstanding precision of $\sim3\cdot$10$^{-8}$ \cite{Kri22}, more than one order of magnitude more accurate than any other element \cite{And99}. The value therefore serves as a calibration standard for negative ion photodetachment measurements and has been used accordingly also in this work to prove the performance of the new application of the FLSR facility.

Low energy electrostatic storage rings, operating in the few 10 keV energy range, such as the CSR at Heidelberg, Germany \cite{vHa16}, DESIREE at Stockholm, Sweden \cite{Tho11}, or RICE at RIKEN, Japan \cite{Nak17}, have qualified as well-suited devices for laser photodetachment experiments on anions \cite{Kri22,Mül21,Nak17}. First high-precision measurements on negative molecules such as OH$^{-}$ have been performed at CSR and DESIREE increasing the fundamental physical understanding of those fragile quantum systems \cite{Mey17,Sch17}. First applications based on molecular anion data such as selectivity enhancement in accelerator mass spectrometry via photodetachment of unwanted isobars were developed recently \cite{Mar22}, To meet the steadily growing interest in spectroscopic studies on anions, the Low-energy Storage Ring of Frankfurt University \cite{Sti10} was equipped with a dedicated plasma discharge ion source followed by a rubidium charge exchange cell to produce anions of gaseous species. Successful storage of negative ions in the FLSR has been shown recently \cite{For20}. Due to its high operation temperature the source produces anions in the ground as well as in metastable low-lying fine structure or excited states.

In this work, we discuss first photodetachment studies on $^{16}$O$^{-}$ anions, stored in the FLSR, induced by radiation from a wide-range tunable high-repetition-rate pulsed titanium-sapphire laser, which was introduced perpendicularly to the ion beam to enable Doppler-free spectroscopy. The three lowest-lying photodetachment channels between the individual fine structure sublevels of the ground states of O$^{-}$ and those of neutral O were measured. Fits using the Wigner curves deliver the electron affinity and the fine structure splitting of the anion ground state, demonstrating the precision achieved and proving the suitability of the FLSR facility for such kind of studies.

\section{Methods}

\subsection{Experimental arrangement and measurement procedure}
\begin{figure}
 \centering
        \includegraphics[pagebox=artbox,width=\textwidth]{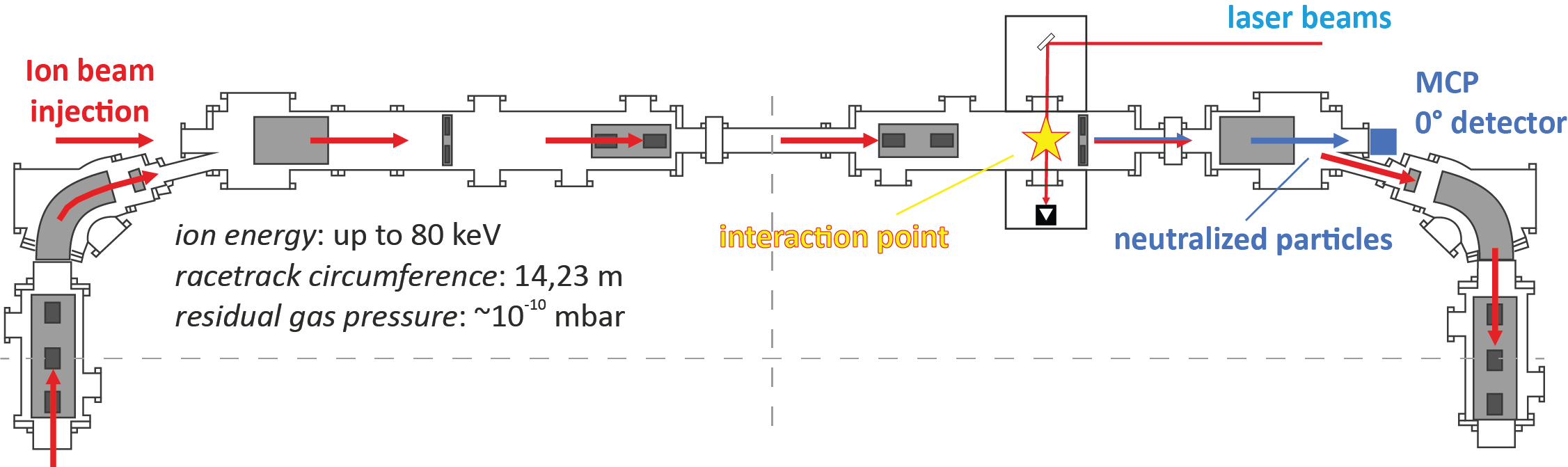}
\caption{Sketch of one long side of the FLSR. Stored ion bunches in injection and circulating are indicated by red arrows, the interaction point between the transversal laser beam and the ion beam is marked by a yellow star and the path of the neutral particles towards the MCP detector is given by a blue arrow. Ion optical elements are depicted in light grey including the quadrupole lenses, deflectors and benders.}
\label{fig1}
\end{figure}

In the present experiment at the FLSR, $^{16}$O$^{-}$ ions are produced from oxygen gas fed into the newly installed RF charge-exchange ion source “Alphatross” from National Electrostatics Corporation (NEC) \cite{NEC,Mid83}. Therein, positive ions are produced by an initial ionization step in a 100\,MHz RF field and a subsequent immediate charge exchange reaction into anions by passing through rubidium vapor. A negative ion beam is extracted from the source and accelerated to a kinetic energy of 20\,keV. Subsequent beam shaping with suitable ion optical elements of the injector beam line, i.e., an Einzel lens, xy deflector plates and further electrostatic quadrupole lenses is followed by mass selection in a magnetic sector field. The ions of interest are subsequently bunched and injected into the storage ring. The race-track geometry of FLSR consists of two long sections for experiments and diagnosis and two short sections connected via 90° deflections, formed by the combination of 15° parallel plate deflectors and 75° cylindrical deflectors. Quadrupole lenses are used to shape the beam. All elements are electrostatic, avoiding mass or velocity selective magnetic deflections \cite{Sti10}. The geometry allows for the detection of neutralized ions created in the long section in the direction of the section axis at 0°. Charge-changing processes towards neutralization can occur either on purpose via laser photodetachment or by unintended influences, e.g., by collisions with residual gas particles (residual gas pressure in the ring $\sim10^{-10}$\,mbar), electric fields or blackbody radiation. All neutral particles, stemming from any of these processes, which the circulating anions could have undergone along the straight section around or at the interaction point, are counted on a position-sensitive microchannel plate (MCP) detector. As visible in Fig.~\ref{fig1}, which sketches one-half of the FLSR including the injection and interaction region, this detector is installed straight after the upcoming bender unit behind the interaction point.
Data acquisition records each individual fast neutral particle hitting the detector together with a high-resolution time stamp and a variety of experimental parameters, i.e., laser power, laser wavelength and all storage ring operation parameters. To perform laser photodetachment studies, the laser wavelength is scanned across the expected photon energy range, yielding the corresponding variation in the neutralization rate as function of photon energy for further detailed numerical analysis, as will be discussed below. Here, the crossed-beam setup was used, which allows to use existing, conventional particle detectors and prevents the first order Doppler effect entirely, directly delivering photodetachment thresholds\footnote{Two geometrical options for laser–ion interaction are in principle possible on the circulating anion ensemble: collinear overlap of laser and ion beam along the straight section of the storage ring or a crossed beam geometry with perpendicular overlap at a given interaction point. For a collinear setup the combination of co- and counterpropagating arrangement allows to eliminate the strong longitudinal Doppler shift of the fast-moving ions by proper data analysis. However, in this case a transparent detector is required with specific properties, which are unaffected by the laser beam transmission (see, e.g., \cite{Han92,War19}).}. Additionally, in this arrangement the use of pulsed lasers implies an excellent discrimination of most of the unselective background from neutrals, produced elsewhere along the beam path, via proper temporal gating of the data readout. The size of the detector is significantly larger compared to the spot produced by the neutralized O$^{-}$ particles impinging on the detector. This allows to further spatially discriminate the signal from background events in the detector by choosing an appropriate region of interest on the detector area. The laser used for the experiment was a titanium-sapphire laser with wide range continuous tunability, applying the tuning via a resonator internal grating, which was designed at the University of Mainz \cite{Tei10}. It is pumped by a high power, high repetition rate frequency doubled Nd:YAG laser (Clark ORC-1000) with 15\,W output power at 7\,kHz repetition rate. The titanium-sapphire laser delivers up to 2\,W power in a spectral range from 1.30 to 1.75\,eV (960-690\,nm) with a linewidth of about 5\,GHz and a pulse duration of around 50\,ns. The laser beam path includes a telescope which is used to optimize the laser beam diameter at the ion beam overlap to maximize the signal on the MCP. Assuming a Gaussian beam profile, a laser beam diameter of 0.5\,mm at FWHM is estimated. The use of a pulsed laser on the circulating ion beam allows to disentangle the laser induced photodetachment signal during laser-on periods from the collisional background during laser-off by suitable temporal gating. During the laser-off periods an average background rate of 15(4) neutralized atoms, coming mainly from collisions with residual gas, was recorded. Permanently recording the power of the laser beam during the spectral scans allows to normalize the neutral particle signal to the photon density in the interaction region, while ion beam current fluctuations are corrected for by analyzing the background intensity during laser off phases. The photon wavelength was recorded using a HighFinesse WS6-600 high precision wavemeter offering an accuracy of 600\,MHz (2.5\,µeV).

\subsection{Data evaluation}

\begin{figure}
 \centering
        \includegraphics[pagebox=artbox,width=0.5\textwidth]{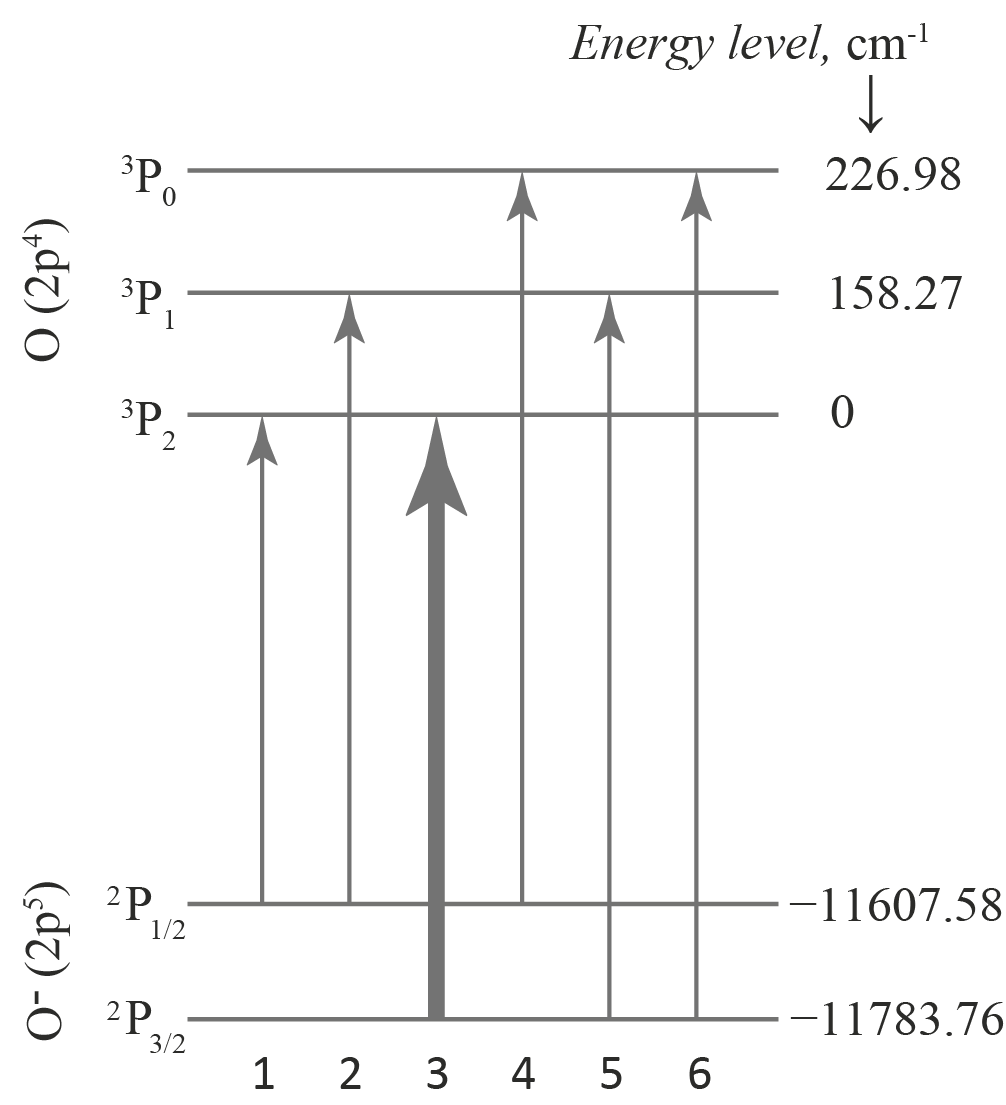}
 \caption{Sketch of the energy levels of O$^{-}$ and neutral O showing the fine-structure. Individual onsets or photodetachment channels between the anion and the neutral are depicted like transitions by arrows and are labelled 1 to 6 in the order of increasing threshold energy. Transition number 3 marked in bold denotes the EA of O$^{-}$ as the value between the two lowest lying levels of anion and neutral, respectively. The energy values are taken from literature \cite{Zin91,Kri22}. The sketch is not to scale.}
\label{fig2}
\end{figure}

The photodetachment pattern of O$^{-}$ at threshold is governed by the fine structure splitting of the ground state of the anion into the two levels 2p$^{5}$ $^{2}$P$^\mathrm{o}_{3/2}$ and 2p$^{5}$ $^{2}$P$^\mathrm{o}_{1/2}$ and the three fine structure levels of the neutral atom 2p$^{4}$ $^{3}$P$_{2,1,0}$. This leads to altogether 6 possible photodetachment channels of Wigner curves with $\Delta$J =1/2, 3/2 and $\Delta$L = 0 as shown in Fig.~\ref{fig2}, causing an up to six-fold overlapping of photodetachment curves above threshold. The energetic splittings of both structures have been precisely investigated already more than 35 years ago. For the anionic ground state configuration, the splitting was reported as 177.13(5)\,cm$^{-1}$, derived as difference of absolute energy values of individual photodetachment steps in electron affinity studies, which were carried out at a collinear-anticollinear laser-ion beam spectrometer with ion beam energy of 3.2\,keV using a tunable infrared dye laser \cite{Neu85}.

For verification of these data and to prove the applicability and accuracy of the FLSR for those kinds of studies, the laser was directed onto a circulating ion ensemble and the energy was scanned multiple times from 1.43 to 1.47\,eV and reverse from 1.47 to 1.43\,eV independently to avoid data acquisition artefacts resulting from the scan direction and a possible delay in data acquisition. In this range the first three photodetachment channels of Fig.~\ref{fig2} are expected and were observed. A typical full range scan is shown in Fig.~\ref{fig3}. Every individual channel corresponds to an outgoing s-electron, for which the development above each threshold is governed by the Wigner law \cite{Wig48}:
\begin{equation}
\sigma \propto ( E_{ph} - E_{0}  ) ^{l+1/2}
\label{Wigner_curve}
\end{equation}
Here, $l$ represents the angular momentum of the outgoing electron, while $E_{ph}$ and $E_{0}$ are the photon energy and the individual threshold energy, respectively. Below the threshold the cross-section for photodetachment is expected to vanish.

The ground state electron affinity of the oxygen anion is obscured by the two lower lying photodetachment curves from the excited state $^{2}$P$_{1/2}^\mathrm{o}$. To obtain the threshold energies, the Wigner law for each channel was fitted to the experimental data, and the resulting fit was used to subtract the lower-lying threshold data of the detachment channels at higher energy. This is repeatedly done until the threshold of interest remained entirely undisturbed in the data, resulting in a more stable numerical description. To account for the linewidth of 5 GHz of the used laser system the Wigner threshold law was convoluted with a corresponding gaussian curve. This convoluted function was then used in the fits of the three thresholds.

Correction for any residual Doppler shift was carried out in the following form: The general relation for the relativistic Doppler effect under an arbitrary angle is given by:
\begin{equation}
f_{r} = \frac{f_{s}}{\gamma ( 1 + \beta \cos \theta_{r} ) }
\end{equation}
where $f_{s}$ is the frequency of the emitted source photon and $f_{r}$ is the frequency seen by the receiver. $\beta$ and $\gamma$ are the relativistic factors and $\theta_{r}$ is the angle between the ion beam and the photon beam. Here, 0° and 180° corresponds to antiparallel and parallel laser propagation with respect to the ion beam, respectively. Due to the crossed-beam geometry between the ions and the laser photons our measurement should only be affected by the relativistic transversal Doppler effect and not by the longitudinal Doppler effect ($\theta_{r}$ = 90°). However, due to technical restrictions during the measurement an angle of 91.0(3)° between the ion and laser beam was set, deviating rather slightly but still significantly from perfect orthogonality. Consequently, a Doppler shift for a kinetic energy of the stored $^{16}$O$^{-}$ ions of 20\,keV is arising with a correction factor of 1.000\,027(8). For quantitative numerical analysis, all raw data were corrected for this Doppler shift and the additional uncertainty was included in the total error budget.

The data analysis and the described fitting to the theoretical description by the Wigner law has been performed using the object-oriented data analysis framework ROOT \cite{ROOT}.

\section{Results And Discussion}

\begin{figure}
 \centering
        \includegraphics[pagebox=artbox,width=\textwidth]{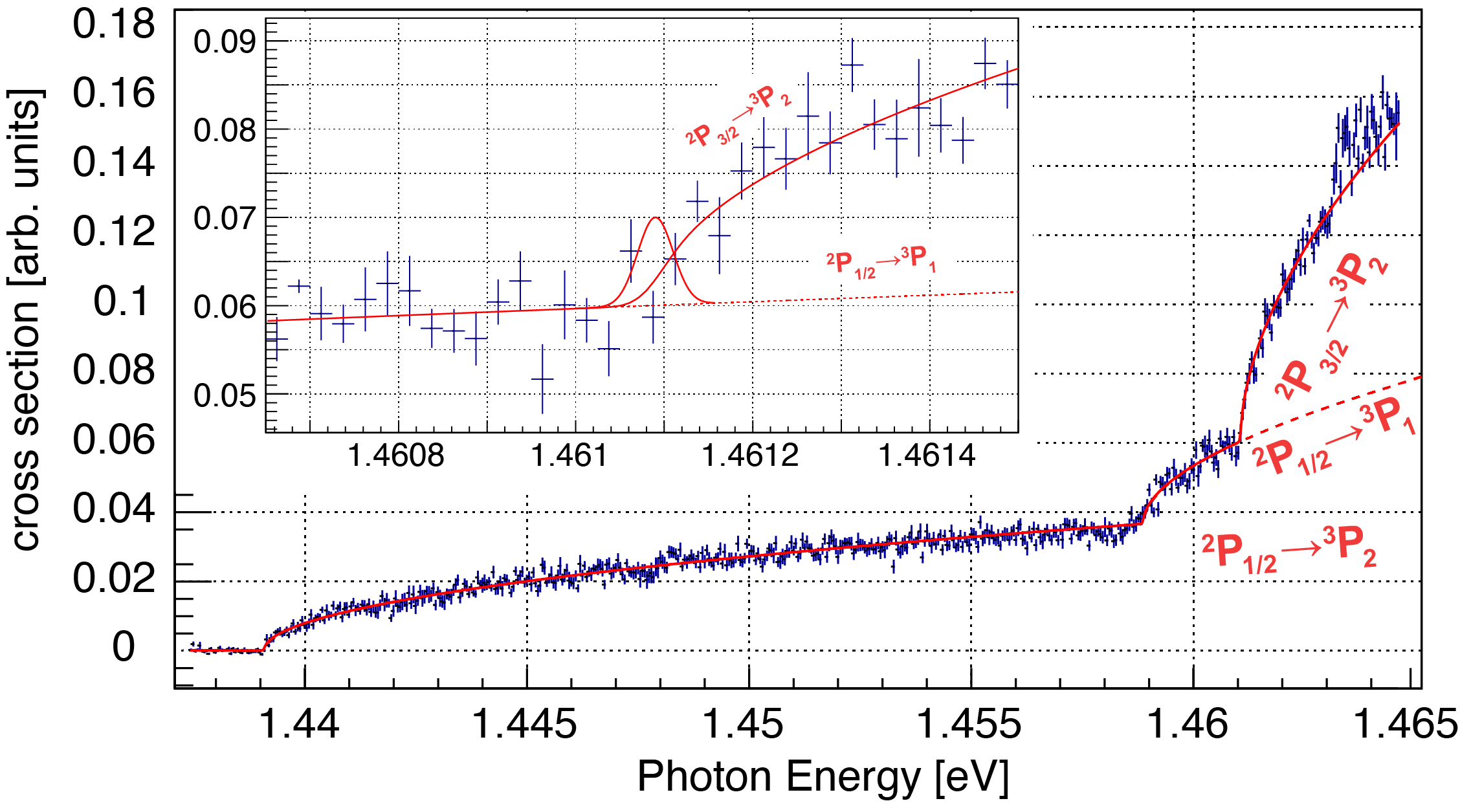}
 \caption{Typical example of an experimental photodetachment spectrum scan in the region of the first three detachment channels of O$^{-}$. The raw data shown in blue are binned using 50\,µeV bins and are not corrected for the Doppler shift. The horizontal error bars show the width of the bins used, while the vertical error bars show the statistical error of the binned data. The dashed lines show the individual thresholds for the photodetachment channels from the O$^{-}$ ($^{2}$P$_{J}^\mathrm{o}$) levels to the O ($^{3}$P$_{J}$) levels. They are labelled according to FIG.~\ref{fig2}. The solid line shows the Wigner law fit of the sum of all three thresholds. The inset shows a zoom into the raw data covering the region  around the EA of O$^{-}$ and binned to 25\,µeV/bin and are also not corrected for the Doppler shift. The horizontal error bars show the width of the bins used, while the vertical error bars show the statistical error of the binned data. The convolution with the bandwidth of the laser (5\,GHz) can be seen at threshold. A gaussian shaped curved is drawn at threshold to show the laser bandwidth.}
\label{fig3}
\end{figure}


A typical full-width spectrum of the uncorrected raw data including all three detachment channels is shown in Fig.~\ref{fig3}. Dotted lines represent the data subtracted from channel 1 and 2, respectively, as described above. Below the first threshold, cross sections slightly below zero are possible due to the statistical count rates in the background and signal time frame. The thresholds fitted with the Wigner law in Eq.~\ref{Wigner_curve} are listed in Tab.~\ref{tab1}. The lowest threshold at 1.439\,133\,(29)\,eV is in agreement with the literature value of 1.439\,150\,(6)\,eV \cite{Neu85}. The threshold for channel 2 is fitted on top of the channel 1 Wigner law curve. The fitted threshold at 1.458\,889\,(49)\,eV is by about 120\,µeV above the literature value of 1.458\,775\,(6)\,eV, as determined 1985 \cite{Neu85}. This discrepancy, which is about twice our accuracy also appears in a milder form in channel 3, i.e. for the EA, where it is about a factor of ten smaller and well lies within our uncertainty of $2\cdot10^{-5}$. Nevertheless the overestimation of both thresholds, regarding especially the much weaker channel 2 hints to an improper fit and extension of the first Wigner curve to the extended range of more than 20\,meV and more, corresponding to almost 200\,cm$^{-1}$. This discrepancy, which has obviously not been seen in \cite{Kri22}, as all lower lying thresholds had been removed by laser cleaning before the detachment study in channel 3, are the topic of a forthcoming publication, which will also include the higher and still weaker detachment channels 4 to 6. The lower red dotted line indicates the influence of channel 1 on the data. It increases slowly with the photon energy. Due to its higher intensity, the EA at 1.461\,122\,(24)\,eV has an increased precision compared to channel 2. A detailed view of the region around the EA is shown in the inset of Fig.~\ref{fig3}, where the convolution of the Wigner law with a Gaussian distribution representing the laser bandwidth is visible. However, while the EA measured in this work is in good agreement with all literature values, the most accurate value of 1.461\,112\,972\,(45)\,eV determined by Kristiansen et al. \cite{Kri22}, surpasses the precision of our work by about three orders of magnitude.

The fine structure splitting in the ion as well as the $^{3}$P$_{1}$ – $^{3}$P$_{2}$ splitting in the atom, as extracted from the data of this work are listed in Tab. 2. The splitting in the anion was calculated as the difference of thresholds in channel 1 and 3 as those are most accurate and in the atom as difference between channel 1 and 2. According to the observed deviations of the fitted positions for different onset curves discussed above, also these values show slightly higher values, than the different high precision literature values included in the table, but deviations for the ion are well within our error and for the atom exceed this only by a factor of 1.5. This precision agrees well with the one of the theoretical predictions given by Godefroid and Froese Fischer \cite{God99}.

\begin{table}
\caption{Fitted thresholds for channel 1 to 3 in Fig.~\ref{fig2}. Literature values from previous experiments are given for comparison. The Doppler shift in first and second order as well as the wavemeter error of 4\,µeV (i.e. 1\,GHz for the used WS6-600 wavemeter at 5\,GHz laser bandwidth) are included.}
\centering
\begin{tabular}{|l|l|l|l|l|l|}
\hline
\multicolumn{2}{|l|}{\multirow{2}{*}{\bfseries Channel}} & \multirow{2}{*}{\bfseries This work (eV)} & \multicolumn{3}{|l|}{\bfseries Literature (Exp.) (eV)} \\
\cline{4-6}
\multicolumn{2}{|l|}{} & & {\bfseries \cite{Neu85} } & {\bfseries \cite{Blo95} } & {\bfseries \cite{Kri22} } \\
\hline
1 & $^{2}$P$^\mathrm{o}_{1/2}$ $\rightarrow$ $^{3}$P$_{2}$ & 1.439\,133\,(29) & 1.439\,150\,(6) & - & 1.439\,157\,53\,(29) \\
\hline
2 & $^{2}$P$^\mathrm{o}_{1/2}$ $\rightarrow$ $^{3}$P$_{1}$ & 1.458\,889\,(49) & 1.458\,775\,(6) & - & - \\
\hline
3 & $^{2}$P$^\mathrm{o}_{3/2}$ $\rightarrow$ $^{3}$P$_{2}$ (EA) & 1.461\,122\,(24) &  1.461\,109\,6(7)\textsuperscript{\textdagger} & 1.461\,110\,7\,(17) & 1.461\,112\,972\,(87) \\
\hline
\multicolumn{6}{l}{{}\textsuperscript{\textdagger}There was an improper conversion from cm$^{-1}$  to eV in \cite{Neu85}.}\\
\end{tabular}
\label{tab1}
\end{table}

\section{Conclusions}
The present experiment on the calibration anion O$^{-}$ demonstrates that the low energy storage ring of Frankfurt University is well suited to study laser photodetachment of negative ions in terms of achievable resolution, precision and efficiency of photon-ion interaction. As a room temperature facility and operating with a negative-ion source capable of populating fine structure levels different spectroscopic studies on molecular and atomic anions are forseen. Using a high-power wide-range tunable Ti sapphire laser first re-measurements of the EA of O$^{-}$ and its fine structure splittings were performed to serve as calibration standard. The results agree with literature values. From fits of the  detachment curves, the electron affinity of $^{16}$O$^{-}$ was derived with a precision of about $3\cdot10^{-5}$.

To evaluate higher-energy detachment channels and cover extended spectral scans around different onsets, careful investigations on the longer-range validity of the simple Wigner law are necessary. Based on our data we have started to consider the zero-core-contribution model introduced by Stehman and Woo \cite{Ste79}. Corresponding detailed analyses are currently ongoing to prepare for further measurements e.g. on the OH$^{-}$ anion and will be published separately.

\begin{acknowledgments}
{O.F. acknowledge support from the German Ministry of Science and Education under Grant No. 05K19SJ1 and 05K22SJ1. V.G., T.N. and K.W. acknowledge support from the German Ministry of Science and Education under Grant No. 05K19UM3 and 05K22UM5.}
\end{acknowledgments}

\begin{table}
\caption{Fine-structure splitting of the O and O$^{-}$ ground state relative to the $^{3}$P$_{2}$ ground state of O. The splittings are calculated as the difference of the threshold energies in Tab.~\ref{tab1}. All values are in cm$^{-1}$.}
\centering
\begin{tabular}{|l|l|l|l|l|}
\hline
\multicolumn{2}{|l|}{\multirow{3}{*}{\bfseries Transition}} & \multicolumn{3}{|l|}{\bfseries Fine structure splitting} \\
\cline{3-5}
\multicolumn{2}{|l|}{} & \multirow{2}{*}{\bfseries This work} & \multicolumn{2}{|c|}{\bfseries Literature} \\
\cline{4-5}
\multicolumn{2}{|l|}{} &  & {\bfseries Experiment} & {\bfseries Theory \cite{God99}} \\
\hline
O & $^{3}$P$_{1}$ - $^{3}$P$_{2}$ & 159.35\,(62) & \begin{tabular}{@{}l@{}}158.268\,741\,(5) \cite{Zin91} \\ 158.302\,98\,(7) \cite{Say79} \end{tabular} & 159.41 \\
\hline
O$^{-}$ & $^{2}$P$^\mathrm{o}_{1/2}$ – $^{3}$P$^\mathrm{o}_{3/2}$ & 177.35\,(42) & \begin{tabular}{@{}l@{}}177.13 (5) \cite{Neu85} \\ 177.084 (14) \cite{Blo01} \\ 177.082 57 (24) \cite{Kri22} \end{tabular} & 178.33 \\
\hline
\end{tabular}
\label{tab2}
\end{table}

%
%



All data supporting the findings of this study are available from the authors on reasonable request.





\begin{thebibliography}{}
\bibitem{And04}	T. Andersen, Atomic negative ions: structure, dynamics and collisions, Phys. Rep. 394, 157-313 (2004)
\bibitem{Bil00}	R. C. Bilodeau, H.K. Haugen, Experimental Studies of Os$^{-}$: Observation of a Bound-Bound Electric Dipole Transition in an Atomic Negative Ion, Phys. Rev. Lett. 85, 534-537 (2000)
\bibitem{Wal11}	C.W. Walter, N.D. Gibson, Y.-G. Li, D.J. Matyas, R.M. Alton, S.E. Lou, R.L. Field, D. Hanstorp, L. Pan, D.R. Beck, Experimental and theoretical study of bound and quasibound states of Ce-, Phys. Rev. A 84 (2011) 032514
\bibitem{Kel14}	A. Kellerbauer, A. Fischer, U. Warring, Measurement of the Zeeman effect in an atomic anion: Prospects for laser cooling of Os$^{-}$, Phys. Rev. A 89 (2014) 043430
\bibitem{Wal14}	C.W. Walter, N.D. Gibson, D.J. Matyas, C. Crocker, K.A. Dungan, B.R. Matola, J. Rohlén, Candidate for Laser Cooling of a Negative Ion: Observations of Bound-Bound Transitions in La$^{-}$, Phys. Rev. Lett. 113 (2014) 063001
\bibitem{Tan19}	R. Tang, R. Si, Z. Fei, X. Fu, Y. Lu, T. Brage, H. Liu, C. Chen, C. Ning, Candidate for Laser Cooling of a Negative Ion: High-Resolution Photoelectron Imaging of Th$^{-}$, Phys. Rev. Lett. 123 (2019) 203002
\bibitem{Tan21}	R. Tang, Y. Lu, H. Liu, C. Ning, Electron affinity of uranium and bound states of opposite parity in its anion, Phys. Rev. A 103 (2021) L050801
\bibitem{Hot75}	H. Hotop, W.C. Lineberger, Binding energies in atomic negative ions, J. Phys. Chem. Ref. Data 4, 539–576 (1975) 
\bibitem{And99}	T. Andersen, H. K. Haugen, H. Hotop; Binding Energies in Atomic Negative Ions: III., J. Phys. Chem. Ref. Data 1 November 1999; 28 (6): 1511-1533. 
\bibitem{Blo95}	C. Blondel, Recent experimental achievements with negative ions, Phys. Scr. T58, 31-42 (1995) 
\bibitem{God99}	M. R. Godefroid, C. Froese Fischer, Isotope shift in the oxygen electron affinity, Phys. Rev. A 60 (1999) R2637
\bibitem{Kri22}	M.K. Kristiansson, K. Chartkunchand, G. Eklund, et al High-precision electron affinity of oxygen, Nat. Comm. 13, 5906 (2022) 
\bibitem{vHa16}	R. von Hahn et al., The cryogenic storage ring CSR, Rev. Sci. Instrum. 87, 063115 (2016)
\bibitem{Tho11}	R. D. Thomas et al. The double electrostatic ion ring experiment: a unique cryogenic electrostatic storage ring for merged ion–beams studies. Rev. Sci. Instrum. 82, 065112 (2011)
\bibitem{Nak17}	Y. Nakano, Y. Enomoto, T. Masunaga, S. Menk, P. Bertier, T. Azuma; Design and commissioning of the RIKEN cryogenic electrostatic ring (RICE), Rev. Sci. Instrum. 88 (2017) 033110
\bibitem{Mül21}	D. Müll, F. Grussie, K. Blaum, S. George, J. Göck, M. Grieser, R. von Hahn, Z. Harman, Á. Kálosi, Metastable states of Si$^{-}$ observed in a cryogenic storage ring, Phys. Rev. A 104 (2021) 032811
\bibitem{Mey17} C. Meyer, A. Becker, K. Blaum, C. Breitenfeldt, S. George, J. Göck, M. Grieser, F. Grussie, E.A. Guerin et al., Radiative Rotational Lifetimes and State-Resolved Relative Detachment Cross Sections from Photodetachment Thermometry of Molecular Anions in a Cryogenic Storage Ring, Phys. Rev. Lett. 119, 023202 (2017)
\bibitem{Sch17} H.T. Schmidt, G. Eklund, K.C. Chartkunchand, E.K. Anderson, M. Kamińska, N. de Ruette, R.D. Thomas, M.K. Kristiansson, M. Gatchell et al., Rotationally Cold OH$^{-}$ Ions in the Cryogenic Electrostatic Ion-Beam Storage Ring DESIREE, Phys. Rev. Lett. 119, 073001 (2017)
\bibitem{Mar22}	M. Martschini, J. Lachner, K. Hain, M. Kern, O. Marchhart, J. Pitters, A. Priller, P. Steier, A. Wiederin, A. Wieser, R. Golser. 5 years of ion-laser interaction mass spectrometry —status and prospects of isobar suppression in AMS by lasers, Radiocarbon., 64(3):555-568 (2022)
\bibitem{Sti10}	K.E. Stiebing, V. Alexandrov, R. Dörner, S. Enz, N.Yu. Kazarinov, T. Kruppi, A. Schempp, H. Schmidt Böcking, M. Völp, P. Ziel, M. Dworak, W. Dilfer, FLSR – The Frankfurt low energy storage ring, Nuc. Instr. and Meth. in Phys. Res. Section A, Volume 614, Issue 1 (2010) 10-16
\bibitem{For20}	O. Forstner, J. Müller, K.E. Stiebing, Opportunities for negative ions studies at the Frankfurt Low-energy Storage Ring (FLSR), Hyp. Int. 241, 53 (2020). 
\bibitem{NEC}	National Electrostatics Corporation (NEC), https://www.pelletron.com/product-category/ion-beam-sources/
\bibitem{Mid83}	R. Middleton, A versatile high intensity negative ion source, Nucl. Instr. Methods 214, 139–150 (1983)
\bibitem{Han92}	D. Hanstorp, A secondary emission detector capable of preventing detection of the photoelectric effect induced by pulsed lasers, Meas. Sci. Technol. 3, 523 (1992)
\bibitem{War19}	J. Warbinek, D. Leimbach, D. Lu, K. Wendt, D.J. Pegg, A. Yurgens, D. Hanstorp, J. Welander, A graphene-based neutral particle detector. Appl. Phys. Lett. 114, 061902 (2019)
\bibitem{Tei10}	A. Teigelhöfer, P. Bricault, O. Chachkova, M. Gillner, J. Lassen, J. P. Lavoie, J. Meißner, W. Neu, K. Wendt, Grating tuned Ti:Sa laser for in-source spectroscopy of Rydberg and autoionizing states, Hyperfine Interact 196, 161–168 (2010)
\bibitem{Neu85}	D.M. Neumark, K.R. Lykke, T. Andersen, W.C. Lineberger, Laser photodetachment measurement of the electron affinity of atomic oxygen, Phys. Rev. A 32 (1985) 1890
\bibitem{Wig48}	E.P. Wigner, On the Behavior of Cross Sections Near Thresholds, Phys. Rev. 73, 1002 (1948)
\bibitem{ROOT}	R. Brun and F. Rademakers, ROOT - An Object Oriented Data Analysis Framework, Proceedings AIHENP'96 Workshop, Lausanne, Sep. 1996, Nucl. Instr. and Meth. A 389, (1997) 81-86, https://root.cern/
\bibitem{Blo01}	C. Blondel, C. Delsart, C. Valli, S. Yiou, M. R. Godefroid, S. Van Eck, Electron affinities of $^{16}$O, $^{17}$O, $^{18}$O, the fine structure of $^{16}$O$^{-}$, and the hyperfine structure of $^{17}$O$^{-}$, Phys. Rev. A 64, 052504 (2001)
\bibitem{Zin91}	L.R. Zink, K.M. Evenson, F. Matsushima, T. Nelis, R.L. Robinson, Atomic Oxygen Fine-Structure Splittings with Tunable Far-Infrared Spectroscopy, Astroph. J. 371, L85-L86 (1991)
\bibitem{Ste79} R.M. Stehman and S.B. Woo, Zero-core-contribution model and its application to photodetachment of atomic negative ions, Phys. Rev. A 20, 281 (1979)
\bibitem{Say79} R.J. Saykally; K.M. Evenson, Laser magnetic resonance measurement of the 2 $^{3}$P$_{2}$–2 $^{3}$P$_{1}$ splitting in atomic oxygen J. Chem. Phys. 71, 1564–1566 (1979)
\end{thebibliography}
\end{document}